\documentclass[12pt]{scrartcl}

\usepackage{graphicx}

\title{Comparison of pulsed electroacoustic and thermally stimulated depolarization current measurements of thermally poled PET electrets}
\date{}
\author{
\large S.E. Parsa, J.C. Ca\~nadas, J.A. Diego, M. Mudarra and J. Sellar\`es\thanks{E--mail: \texttt{jordi.sellares@upc.edu}} \\
\large Departament de F\'{\i}sica, Universitat Polit\`ecnica de Catalunya\\
\large Campus de Terrassa, c. Colom~1, E-08222 Terrassa, Spain.
}

\begin{document}

\maketitle

\begin{center}
\textbf{Abstract}
\end{center}

\begin{abstract}

\small

We have compared measurements of a set of polyethylene terephthalate (PET) electret samples by means of pulsed electroacoustic method (PEA) and thermally stimulated depolarization current (TSDC) techniques. Experimental parameters such as the combined thermal and electrical history and the electrode type have been selected in order to correlate the polarization mechanisms revealed by TSDC with the charge profile measured by PEA in five different cases. Existing deconvolution procedures for PEA have been improved as a means to enhance the calibration of PEA signals in the case of thin samples. Samples where the $\alpha$ dipolar relaxation or the $\rho$ space charge relaxation is activated show a uniform polarization that manifests itself as image charge at the electrodes. In the experiments where external charge carriers are injected into the sample, the same poling procedure has been tested under different electrode configurations. Charge profiles are qualitatively similar in all of them but the depolarization currents show clearly different behavior. These differences are explained, on the one hand, by the different blocking behavior of vacuum-deposited aluminum electrodes with regards to electrodes with a thin air gap and, on the other hand, by the distinct behavior of electrodes with an air gap for both directions of the charge carriers. Numerical analysis of the polarization of TSDC peaks and charge per unit area of charge profiles supports this interpretation and confirms the relationship between both measurement techniques. All in all, PEA in combination with TSDC turns out to be a useful technique in the study of thermally poled electrets, either in the study of relaxations or of external charge.


\end{abstract}

\section{Introduction}

An electret is made of a dielectric material that has been poled in such a way that it creates quasi--permanent external and internal electric fields \cite{eguchi25, gerhard87}. Electrets can be poled by several procedures such as corona charging, electron beam irradiation or thermal poling, among other ones \cite{gerhard87}. Thermal poling (TP) is a well-known technique to obtain electrets. It is the basis of spectroscopic techniques such as thermally stimulated depolarization current (TSDC) \cite{chen81}, that has been used for decades to study dipolar and space charge relaxation in solids. For this reason, the polarization that can be obtained by TP has been extensively studied from a macroscopic point of view \cite{vanturnhout99}. In spite of this fact, there is still information about the microscopic mechanisms that are activated by TP that is not fully determined yet.

For example, it seems out of question that the $\alpha$ relaxation is the dielectric signature of the structural relaxation \cite{belana85} and that it is due to the cooperative reorientation of molecular dipoles and that the $\rho$ relaxation is due to free charge trapped in localized states \cite{mudarra97}. These relaxations can be detected, for instance, as peaks in a TSDC spectrum. Nevertheless, it is widely accepted that additional mechanisms are needed to fully explain these phenomena. A throughout understanding of the mechanisms that give rise to a relaxation is especially useful when interpreting data from spectroscopic techniques and it can also give clues on how to obtain more stable electrets or electrets that are better suited for particular applications.

One important piece of information related to the microscopic mechanisms is the charge profile of the electret \cite{montanari05}. Many studies have been made about the charge profile of corona charged \cite{ono04} or electron beam irradiated electrets \cite{maeno88, banda18} but there are fewer about thermally poled electrets \cite{kazansky96, hoang14}. This is because they tend to store less charge and to be less stable than electrets poled by other means. Also, polarization is more uniform which makes it more difficult to detect the induced charge because of its proximity to the electrodes. As a consequence, their charge profile is more difficult to measure and results are less conclusive.

Several methods can be used to measure the charge profile of an electret sample \cite{lewiner05} up to a certain resolution \cite{hole08}. Acoustic methods can give the charge or field distribution in the sample. Additional quantities, such as surface potential, can be obtained by calculation from these ones \cite{chen07}. Depending on the signal--generation process, acoustic methods can be divided into LIPP (laser-induced pressure pulse), PPS (piezoelectrically-generated pressure step) or PEA (pulsed electroacoustic) methods. In all these cases, pressure waves generated  either at the sample surface or at charge layers in the bulk, propagate through the sample with the speed of sound waves. The deformations that are produced in the sample cause currents or voltages at the electrodes due to charge displacement, changes in electric permittivity or because of the piezoelectric effect at a transducer \cite{sessler97}.

Another important information about an electret is its relaxational behavior. It is not only important to know where its charge or its polarization is located but also how these quantities relax to their equilibrium values. TSDC is a well-proven technique that can be used successfully to characterize these relaxations. One of the characteristics that can be obtained with TSDC is the polarization attained when a relaxation is activated. This one will be interesting for our study because it can be cross-checked with the charge profile results. 

The PEA method is based on the electrostatic force. An externally applied electric field pulse induces a perturbing force on the charges that are present in the material. This perturbation generates a sound wave which originates from the charge distribution. The acoustic signal is detected by a piezoelectric transducer placed on one of the electrodes \cite{maeno88}. The space-charge profile information contained in the signal is measured and calibrated through the use of digital signal processing \cite{chen06}. The main difference in contrast to other methods, such as PPS or LIPP, where the pressure wave is generated externally, is that the acoustic wave is generated internally by the space charge \cite{ahmed97}. Initially, deconvolution or other data processing was essential to obtain the charge profile. Some very important improvements were made to the transducers which resulted in a more accurate signal and lessened the need for deconvolution \cite{li94}. The duration of the voltage pulse in PEA measurements is usually between 5 to 40~ns and determines the spatial resolution of the measurements \cite{fleming05}. Nowadays, there is a renewed interest in PEA because of its reliability to detect space charge accumulation in the dielectric insulation of high-voltage direct-current cables \cite{rizzo19}.

The aim of this work is to develop new ways to improve our knowledge about thermally poled electrets. In particular, we want to know if acoustic pulse methods can be used to gain a better understanding of the mechanisms that give rise to polarization in thermally poled electrets. To this end, we will measure the charge profile of samples prepared in such a way so that different mechanisms are activated and we will compare this charge profile with the TSDC relaxational spectra of the samples. We expect to find out the spatial distribution of the mechanisms that give rise to dipolar and space charge mechanisms and the role of external charge, which can depend on  the electrode type. We will adapt existing deconvolution procedures for PEA so that they are better suited for thin samples. The comparison of charge per unit area as measured by PEA and the polarization of the sample as measured by TSDC will allow us to validate our assumptions.

\section{Material and methods}

Experiments have been performed on samples of amorphous (as-received) polyethylene terephthalate (PET). PET was supplied by Autobar Packaging~SA in the form of sheets with a thickness of $320$~$\mu$m. 
Characterization by differential scanning calorimetry (DSC) has shown that the glass transition takes place at $80$~$^\circ$C and cold crystallization begins around $100$~$^\circ$C. It has also been checked by DSC that as-received material crystallinity degree is below our detection threshold and, therefore, is less than $3$\% \cite{sellares10}.

Samples of $2.5 \times 2.5$~cm$^2$ were cut and aluminum electrodes with a diameter of $2$~cm were vacuum deposited on one side or on both sides of the samples. 

\begin{figure}
\begin{center}
\includegraphics[width=8cm]{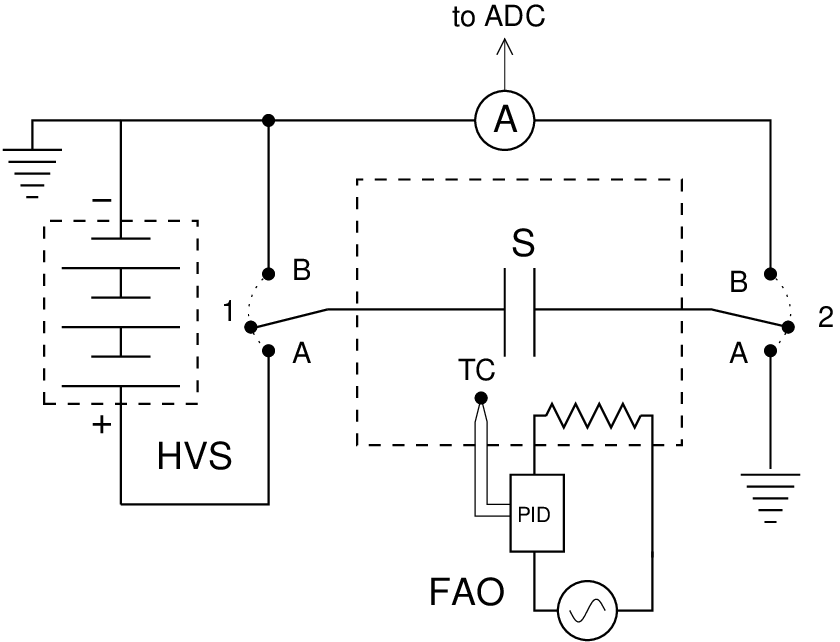}
\caption{Scheme of a TSDC setup.}\label{tsdc-setup}
\end{center}
\end{figure}

Prior to charge profile measurements, samples were thermally poled using a TSDC setup. A scheme of a TSDC setup is presented in Figure~\ref{tsdc-setup}.
The sample (S) is poled by the high voltage source (HVS) setting the two--way switches~1 and~2 in position~A. Once poled, switches are commuted to position~B to depole the sample and record the depolarization current through the amperimeter. A sample holder ensures contact of the sample with the setup electrodes. The sample holder is located inside a forced air oven (FAO) driven by a PID controller with a thermocouple (TC) as input.  

\begin{figure}
\begin{center}
\includegraphics[width=12cm]{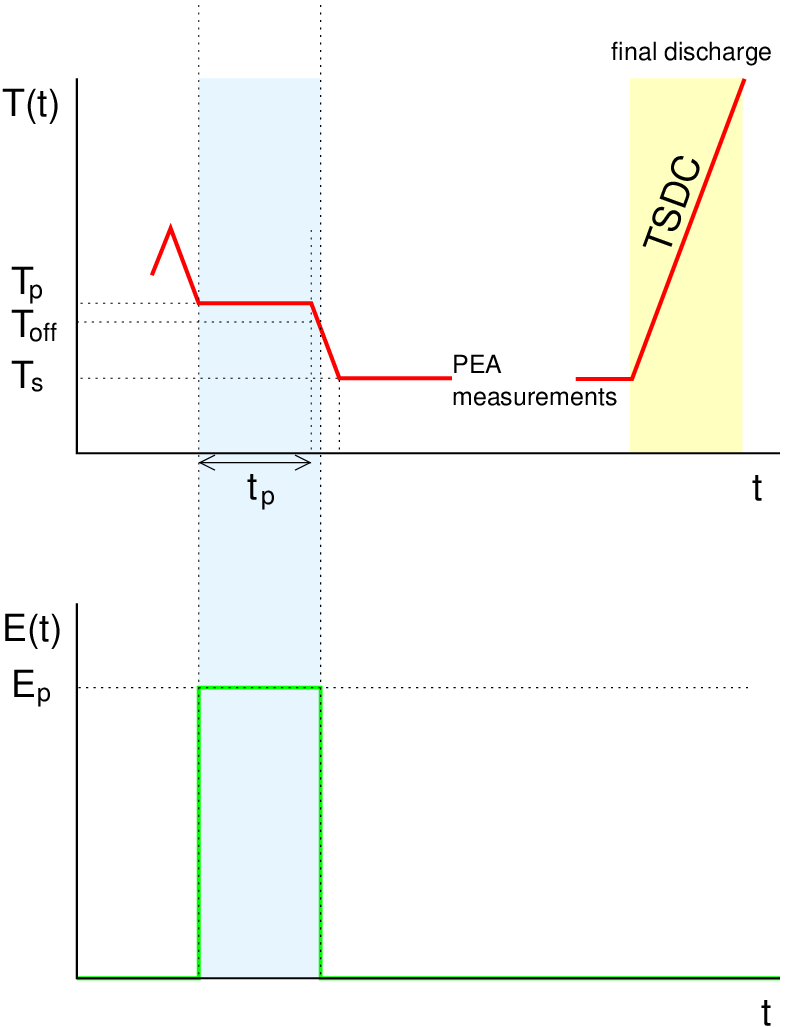}
\caption{Thermal and Electrical history of the sample in the experiments that have been performed.}\label{thermal-poling2}
\end{center}
\end{figure}

The combined thermal and electrical history in our experiments is described in Figure~\ref{thermal-poling2}.
The sets of experimental parameters (A--E) that have been employed are summarized in Table~\ref{exp_parameters}.
\begin{table}
\caption{Experimental parameters employed. The heating and the cooling rates are $2$~$^\circ$C/min in all the non-isothermal stages.}
\label{exp_parameters}
\begin{center}
\begin{tabular}{ccccccc}
Experiment & $Al$ electrode side & $T_0$~($^\circ$C) & $T_{p}$~($^\circ$C)           & $t_{p}$~(s) & $T_{off}$~($^\circ$C) & $T_{s}$~($^\circ$C) \\ \hline
$A$        & Both                & 95                & 65                            & 600         & 55                    & 25                  \\
$A(T_p)$   & Both                & 95                & 55, 50, 45, 35                & 600         & $T_p - 10$            & 25                  \\
$B$        & Both                & 140               & 95                            & 0           & 80                    & 25                  \\
$C$        & Both                & 95                & 80                            & 1200        & 75                    & 25                  \\
$D$        & Negative            & 95                & 80                            & 1200        & 75                    & 25                  \\
$E$        & Positive            & 95                & 80                            & 1200        & 75                    & 25                  \\
\end{tabular}
\end{center}
\end{table}

The sample is initially heated up to $95$~$^\circ$C in all experiments except~B, a temperature well above the glass transition but not high enough to crystallize the sample significantly. In this way, the thermal history of the sample is erased and the experiment begins in a structural equilibrium state. In experiment~B, instead, the initial temperature is $140$~$^\circ$C in order to crystallize the sample. This is done to suppress as much as possible the dipolar relaxation, that takes place in the amorphous phase.

After this initial heating, the sample is cooled down to the poling temperature. Samples A, C, D and E have an isothermal poling stage while sample B is poled non--isothermally. In all cases, the sample is poled by applying a voltage of $5000$~V. The temperature at which the field is switched off is selected to put an upper limit on the frequency of the relaxations that are activated. After poling, the sample is cooled down to room temperature (storage temperature) at a controlled rate. Once at room temperature, polarization is relatively stable so procedures for measuring the charge profile can be performed. Finally, the sample is discharged by means of TSDC.

\begin{figure}
\begin{center}
\includegraphics[width=12cm]{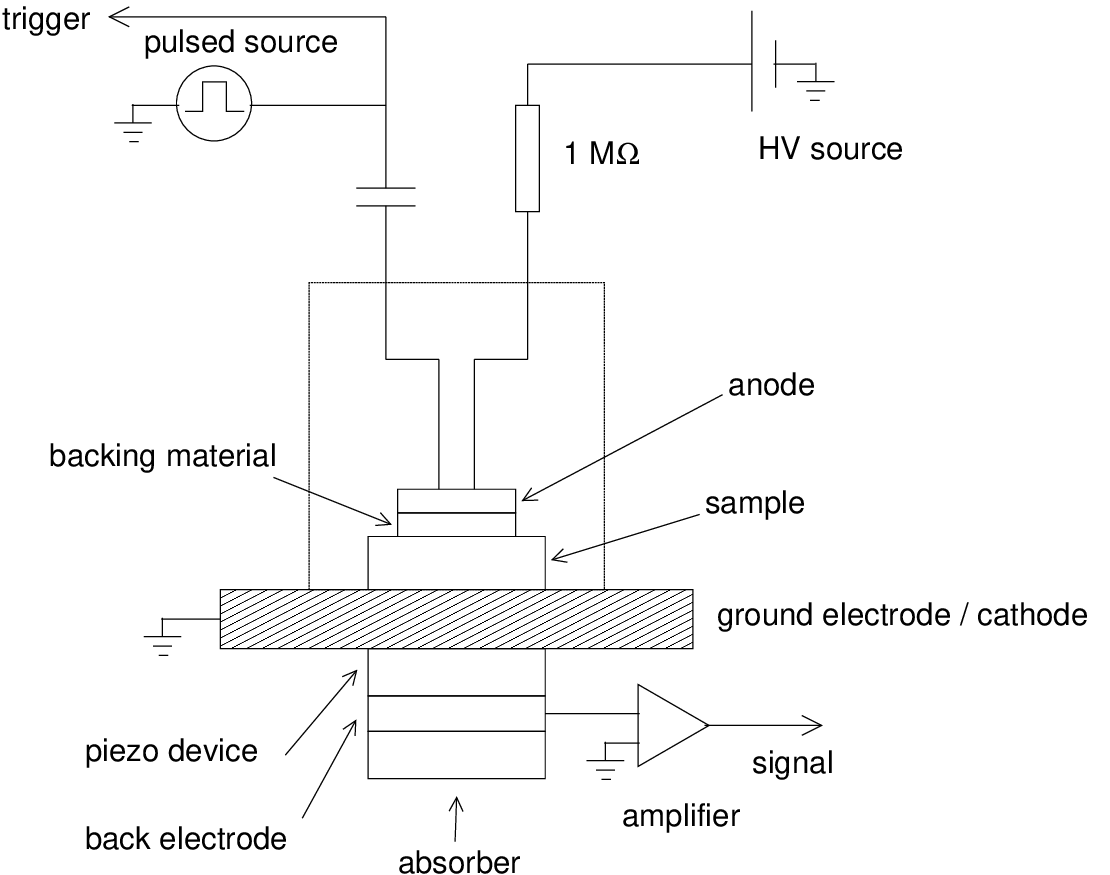}
\caption{Scheme of the PEA setup}\label{pea-setup}
\end{center}
\end{figure}

The charge profile has been measured by the PEA method. A commercial setup provided by TechImp (Italy) was employed. A typical PEA setup is presented in Figure~\ref{pea-setup}. 
Within this method, an electrical pulse (in our setup with a width of $20$~ns and an amplitude of $400$~$V$) is applied to the sample. This electrical pulse produces a back-and-forth movement of the charges of the system that produces an acoustic signal given by \cite{maeno88}
\begin{equation}
p(t) = C v \int_0^t \rho(v \tau) e(t - t_0 - \tau) \, \mathrm{d}\tau
\label{pea_int}
\end{equation}
where $v$ is the speed of sound in the material, $e(t)$ is the electric field of the electrical pulse, $t_0$ is a time offset due to the electrode thickness, $\rho$ is the charge density profile and $C$ is an adimensional constant that depends on the acoustic properties of the material and can be expected to be $0.5$ for homogeneous materials in a unidimensional waveguide~\cite{li94}. $\rho$ will correspond to the charge profile across the whole setup, not only at the sample but also at the electrodes if transmission of the sound waves across the interfaces between the sample and the electrodes is complete. To increase transmission as much as possible air is substituted at the interfaces by silicon oil.

This acoustic signal $p(t)$ is registered by a piezoelectric sensor, amplified and captured by an oscilloscope. For sharp electrical pulses, $p(t)$ is approximately proportional to $\rho(vt)$.

During measurements, a direct current high voltage can be applied to the sample to pole it in--place but we have used this possibility only in calibration measurements. In all the other charge profile measurements the sample is poled previously in the TSDC setup.

\section{Calculation}

The time duration of the electrical pulse in our setup is short enough so that an acceptable resolution can be obtained without deconvolution. Nevertheless, the deconvolution process not only accounts for the shape of the electrical pulse but also serves as a calibration that allows us to convert the voltage measured by the oscilloscope to proper charge density units and to deal with the characteristic response of the sensor. In this section we describe how we perform the deconvolution and calibration of PEA data with a digital signal processing method adapted from reference \cite{arnaout16}, because the commercial software supplied by TechImp gave poor results on thin samples. 

From the discretization of Equation~\ref{pea_int}, it follows that the signal captured by the oscilloscope and sampled as a vector $V$ can be expressed as the product of a convolution matrix $H_\mathrm{PEA}$ and a vector that represents the charge profile $\rho$. 
\begin{equation}
V = H_\mathrm{PEA} \rho
\label{matrices}
\end{equation}
Moreover, since $H_\mathrm{PEA}$ represents a convolution it is a Toeplitz matrix and the product is commutative.
\begin{equation}
V = \Gamma h_\mathrm{PEA}.
\label{matrices2}
\end{equation}
The Toeplitz matrices $H_\mathrm{PEA}$ and $\Gamma$ can be easily determined from vectors $h_\mathrm{PEA}$ and  $\rho$, respectively using publicly available routines for circulant matrices \cite{scipy}. 

To calibrate a given PEA experiment a particular signal $V_\mathrm{cal}$ has to be obtained in a calibration experiment where the corresponding $\rho_\mathrm{cal}$ can be calculated theoretically. We measure a discharged sample with the same characteristics and thermal history as the one employed in the experiment. An electric field of $E_\mathrm{pol} = 5$~kV/mm is applied in order to avoid charge injection and measure $V_\mathrm{cal}$ in a situation that can be considered as an ideal capacitor. For this reason, the charge profile will only consist of the two image charges at the electrodes
\begin{equation}
\rho_\mathrm{cal}(x) = \epsilon_0 E_\mathrm{pol} \left\{ \delta(x-l) - \delta(x) \right\}
\label{rho_cal}
\end{equation}
where $x=0$ is the position of the ground electrode and $l$ is the thickness of the sample. Deliberately, we omit the additional image charge that would result from taking into account the relative permittivity $\epsilon_r \approx 3.3$ \cite{lide04} of the sample because it is canceled by the induced charge that appears at the internal face of the electrodes. Therefore, Equation~\ref{rho_cal} assumes a complete transmission of sound waves at the sample-electrode interfaces.

The key point of the method is that for the discrete version of $\rho_\mathrm{cal}$ we will use gaussian functions instead of deltas of Dirac
\begin{equation}
\rho_\mathrm{cal} = \epsilon_0 \sqrt{\frac{\pi}{2}}E_\mathrm{pol} \left\{ \exp(-\frac{2(x-l)^2}{d^2}) - \exp(-\frac{2x^2}{d^2}) \right\}
\label{rho_cal2}
\end{equation}
with no change on the total charge at the electrodes. Now $x$ is a discretized representation of the position variable. To represent it so that it matches $V_\mathrm{cal}$ we need to know the speed of sound $v$ in the sample, the sampling rate of $V_\mathrm{cal}$ and the index of the element of $V_\mathrm{cal}$ that corresponds to the electrode at $x=0$. This information is obtained from the sampling rate of the oscilloscope and an analysis of the maxima and minima of $V_\mathrm{cal}$. 

\begin{figure}
\begin{center}
\includegraphics[width=12cm]{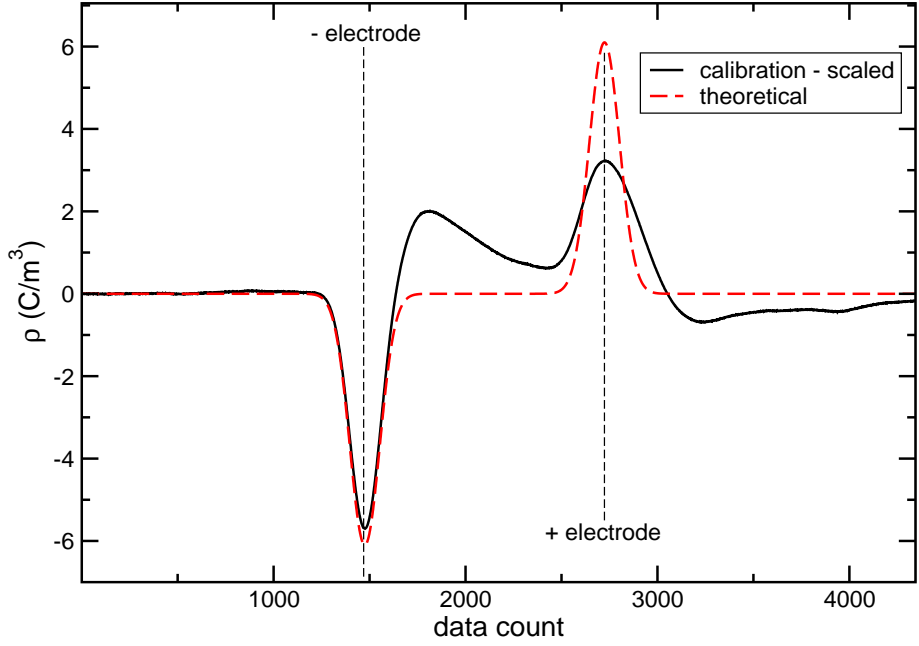}
\caption{PEA measurement of the charge profile of a PET sample while $5$~kV/mm are being applied at room temperature}\label{deconv}
\end{center}
\end{figure}

The $d$ parameter has dimensions of length and plays a role similar to the Wiener and Gaussian filter parameters in the original deconvolution scheme \cite{maeno88}. It should be small compared to the thickness of the sample but large enough to obtain artifact--free results. In most cases, $d = 30$~$\mu$m is an appropriate choice for our signals. An example of an experimental $V_\mathrm{cal}$ calibration signal and its calculated $\rho_\mathrm{cal}$ charge profile is presented in Figure~\ref{deconv}.

We can obtain $h_\mathrm{PEA}$ by inversion of Equation~\ref{matrices2} and applying it to the calibration case
\begin{equation}
h_\mathrm{PEA} = \Gamma_\mathrm{cal}^{-1} V_\mathrm{cal}
\label{matrices3}
\end{equation}
and finally we obtain the deconvoluted charge profile $\rho$ from $V$ by inversion of Equation~\ref{matrices}
\begin{equation}
\rho = H_\mathrm{PEA}^{-1} V
\label{matrices4}
\end{equation}
where $H_\mathrm{PEA}$ can be specified in terms of $h_\mathrm{PEA}$.

\begin{figure}
\begin{center}
\includegraphics[width=12cm]{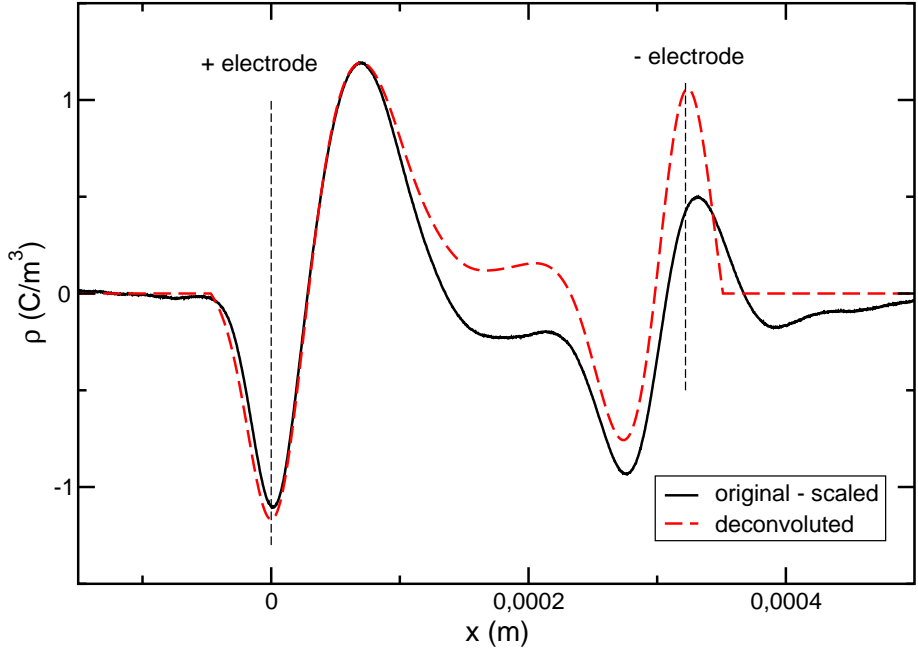}
\caption{Charge profile of a PEA experiment: original signal (continuous line, arbitrary units), deconvoluted signal (dashed line)}\label{deconv2}
\end{center}
\end{figure}

A comparison between a charge profile obtained using this method and the original signal recorded by the oscilloscope is presented in Figure~\ref{deconv2}.
To ease the comparison between the two curves, the original curve has been scaled in such a way that there is maximum likelihood in the image charge at the anode (positive electrode). Aside from the calibration, changes in the shape of the curve are rather limited, which is logical when using short pulses, but are significant at the cathode (negative electrode) where a much sharper image charge is obtained.

The difference of the extreme values of the peaks located at the electrodes in the experimental signal indicates that our assumption of complete transmission at the interfaces is not exact even though we expect it to be good enough to obtain meaningful results.

\section{Results}

Experiments $A$, $B$ and $C$ are performed on samples that have vacuum-deposited aluminum electrodes on both sides of the sample, and differences in the results will be due to different thermal and electrical histories. In experiment $A$ the thermal history has been chosen to activate the $\alpha$ relaxation and the sample remains in amorphous state throughout the experiment. 

Instead, sample $B$ is brought to a semi--crystalline state at the beginning of the experiment. This is done to suppress as much as possible the $\alpha$ relaxation that takes place in the amorphous phase. The non--isothermal poling stage and the range of temperatures at which the electric field is activated is meant to activate the $\rho_c$ relaxation, which is physically similar to the $\rho$ relaxation but takes place in the crystalline phase~\cite{colomer98}. 

Finally, the experimental parameters of experiment $C$ try to activate a varied set of mechanisms through a long isothermal poling stage at a temperature close to $T_g$. 

\begin{figure}
\begin{center}
\includegraphics[width=12cm]{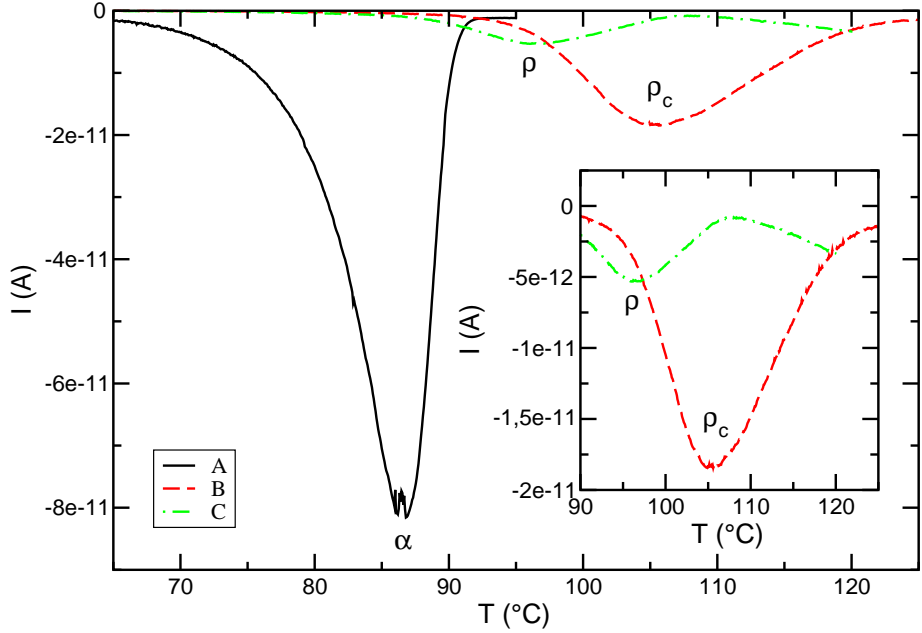}
\caption{TSDC spectra with M2 samples}\label{TSDC-relax}
\end{center}
\end{figure}

The relaxations that are activated in each experiment can be analyzed by TSDC. The resulting depolarization currents are plotted in Figure~\ref{TSDC-relax}.
In the presented TSDC plots, no PEA measurements are carried on the sample before depolarization so as to register the activated relaxations in the most reproducible way.

The spectrum recorded in the $A$ experiment is dominated by the $\alpha$ relaxation, which has its maximum at $86$~$^\circ$C. In experiment $B$ only one relaxation can be observed, with a maximum at $105$~$^\circ$C. We identify this relaxation as $\rho_c$ because the sample is in semi-crystalline state during the poling stage. Finally, in experiment $C$, the $\rho$ peak appears, located at $97$~$^\circ$C.

\begin{figure}
\begin{center}
\includegraphics[width=12cm]{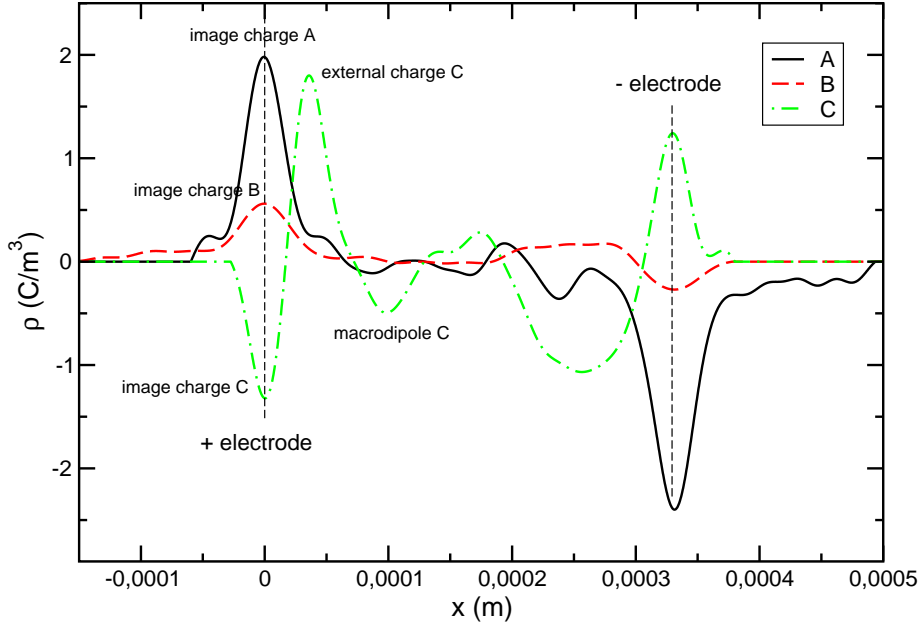}
\caption{Charge profiles of M2 samples obtained with PEA}\label{PEA-relax}
\end{center}
\end{figure}

The charge profile of these samples as measured by PEA is presented in Figure~\ref{PEA-relax}.

In the $A$ experiment the image charge at the electrodes is clearly visible. This charge shields the polarization of the $\alpha$ relaxation, that appears to be uniform since no variation along the bulk of the sample can be observed. 

The charge profile of experiment $B$ corresponds to a sample where $\rho_c$ is activated. As in the previous case, the profile also consists of image charge but with lower charge density. This can be attributed to a lower polarization of the sample, as it can be deduced from the area of the peaks in Figure~\ref{TSDC-relax}. Since polarization appears to be uniform and no accumulation of charge can be observed anywhere in the bulk, we relate the $\rho_c$ relaxation to microscopic displacement of charge carriers (electrons or vacancies) and their subsequent trapping in localized states (charge traps).

Because of the long isothermal poling stage at a temperature close to $T_g$, there is charge injection in experiment $C$. It can be inferred that during the poling stage the aluminum electrodes are non--blocking since they allow charge carriers, either electrons or vacancies, to be injected into the bulk of sample. The charge profile is conditioned by homocharge (charge of the same sign as the closest electrode) due to the external charge of either sign that can be observed in the bulk close to both electrodes. Between the external charge there is a heteropolar macroscopic dipole but with lower density than injected charge. In spite of this fact, the depolarization current in this experiment consists only in heterocurrent. 

\begin{figure}
\begin{center}
\includegraphics[width=12cm]{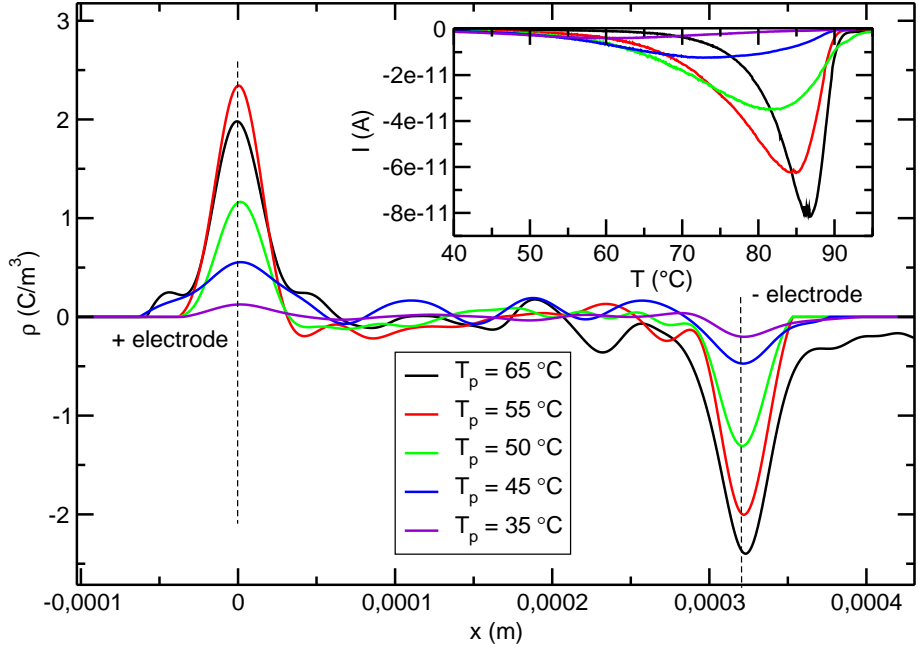}
\caption{Charge profiles of samples with $\alpha$ relaxation activated obtained with PEA. TSDC spectra presented in the inlet}\label{RMA}
\end{center}
\end{figure}

To confirm that the charge distributions observed in the charge profile of experiment $A$ really correspond to image charge and discard that it is due to external charge, we repeated this experiment with several other poling temperatures and labeled them as $A(T_p)$. Results are presented in Figure~\ref{RMA}
In the inlet, TSDC results are plotted. As the poling temperature decreases a different range of mechanisms are activated but the maximum intensity of the $\alpha$ relaxation diminishes. This allows us to study by PEA the charge profile of the samples. These results are presented in the main part of the plot and show an overall trend to decrease for lower poling temperatures. 

It is also interesting to consider the role of the electrodes comparing samples with the same thermal and electrical history and different electrode configuration. This is done through the comparison of experiments $C$, $D$ and $E$. Experiments $D$ and $E$ are performed with samples with only one vacuum-deposited aluminum electrode. This leaves a very thin air gap between the setup electrode and the sample on the other side. During the poling stage, the air gap is on the positive electrode in experiment $D$ and on the negative electrode in experiment $E$.

\begin{figure}
\begin{center}
\includegraphics[width=12cm]{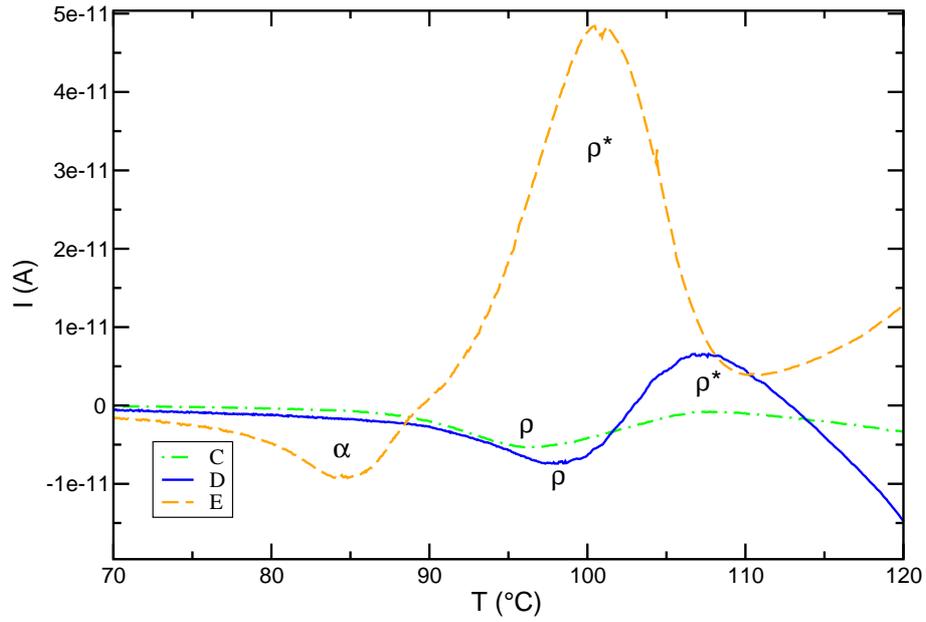}
\caption{TSDC spectra for experiments with the same history and different electrode configuration}\label{TSDC-80C}
\end{center}
\end{figure}

The relaxations that are activated can be obtained from the depolarization currents plotted in Figure~\ref{TSDC-80C}.
What stands out most is that the relaxational spectra of samples with an air gap is more complex. In experiment $D$ we can see the same heterocurrent $\rho$ peak as in experiment $C$, at approximately the same temperature of $97$~$^\circ$C but also a homocurrent peak, labeled as $\rho^*$, with a maximum at $107$~$^\circ$C. In the discharge current of experiment $E$ a huge homocurrent $\rho^*$ peak with a maximum at $101$~$^\circ$C prevails. There is also an heterocurrent peak at $85$~$^\circ$C,  labeled as $\alpha$ because it is probably due to indirect dipolar polarization caused by the homopolar charge. As in previous experiments, the $\rho$ peak corresponds to the relaxation space charge trapped after microscopic displacement. We attribute the $\rho^*$ peak to recombination of external homocharge in the bulk after macroscopic displacement since it is an homocurrent peak and therefore we assume that it is more related to the mobility of external charge carriers.

\begin{figure}
\begin{center}
\includegraphics[width=12cm]{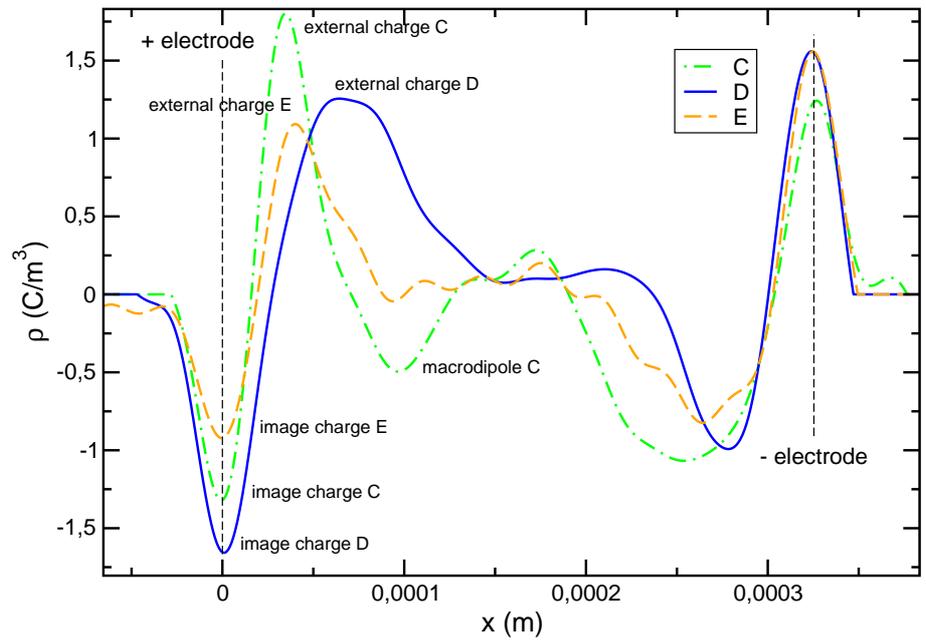}
\caption{Charge profiles for experiments with the same history and different electrode configuration obtained with PEA}\label{PEA-80C}
\end{center}
\end{figure}

The PEA charge profiles corresponding to these experiments are presented in Figure~\ref{PEA-80C}.
These profiles can indicate if the mechanisms registered by TSDC are distributed uniformly across the thickness of the sample or are located in particular regions.

It can be observed in experiment $D$ that the air gap facilitates the penetration of external charge from the positive electrode into the bulk of the sample. This is probably due to a corona charging current made of positive ions in the air gap while poling the sample. Once on the surface of the sample, the positive ions attract electrons leaving vacancies in the bulk. On its part, negative injection is similar in all the three cases. This is a sign that when a strong electric field is applied neither the vacuum-deposited electrodes nor electrodes with an air gap are able to block effectively injection of charge carriers in the sample. 

A numerical study of the polarization that has been attained in the TSDC experiments can be done integrating the depolarization current corresponding to a given relaxation on the time variable. The data obtained is presented in Tables~\ref{polarization} and~\ref{polarization2}.
\begin{table}
\caption{Polarization relaxed in TSDC experiments.}
\label{polarization}
\begin{center}
\begin{tabular}{cccc}
Experiment & $\alpha$ ($\mu$C/m$^2$) & $\rho$ or $\rho_c$ ($\mu$C/m$^2$) & $\rho^*$ ($\mu$C/m$^2$)    \\ \hline
A          & $-64.1$                 &                                   &                            \\
B          &                         & $-28.2$                           &                            \\
C          &                         & $-7.27$                           &                            \\
D          &                         & $-9.39$                           & $4.23$                     \\
E          & $-10.9$                 &                                   & $44.8$                     \\
\end{tabular}
\end{center}
\end{table}
\begin{table}
\caption{Polarization relaxed from the $\alpha$ peak in TSDC experiments.}
\label{polarization2}
\begin{center}
\begin{tabular}{c|ccccc}
Experiment                       & $A$       & $A(55^\circ \mathrm C)$ & $A(50^\circ \mathrm C)$ & $A(45^\circ \mathrm C)$ & $A(35^\circ \mathrm C)$ \\ \hline
$\alpha$ ($\mu$C/m$^2$)          & $-64.1$   & $-84.4$                 & $-66.5$                 & $-32.9$                 & $-11.8$                 \\
\end{tabular}
\end{center}
\end{table}

These data can be compared with the charge per unit area of the charge distributions in the profiles that are presented in Figures~\ref{PEA-relax}, \ref{RMA} and~\ref{PEA-80C}. The charge per unit area corresponding to each distribution of the charge profile is obtained integrating its charge profile along the depth axis. The data obtained is presented in Tables~\ref{sigma_density} and~\ref{sigma_density2}.
\begin{table}
\caption{Charge per unit obtained from PEA measurements.}
\label{sigma_density}
\begin{center}
\begin{tabular}{cccc}
Experiment & img. charge                 & ext. charge           & macr. dipole                 \\
           & ($\mu$C/m$^2$)              & ($\mu$C/m$^2$)        & ($\mu$C/m$^2$)               \\ \hline
A          & $77.2$                      &                       &                              \\
B          & $40.3$                      &                       &                              \\
C          & $-34.5$                     & $53.0$                & $-17.7$                      \\
D          & $-55.5$                     & $89.7$                &                              \\
E          & $-35.2$                     & $40.6$                &                              \\
\end{tabular}
\end{center}
\end{table}
\begin{table}
\caption{Image charge per unit area obtained from PEA measurements with $\alpha$ relaxation activated.}
\label{sigma_density2}
\begin{center}
\begin{tabular}{c|ccccc}
Experiment                          & $A$       & $A(55^\circ \mathrm C)$ & $A(50^\circ \mathrm C)$ & $A(45^\circ \mathrm C)$ & $A(35^\circ \mathrm C)$ \\ \hline
Img. charge ($\mu$C/m$^2$)          & $77.2$    & $81.6$                  & $40.4$                  & $29.3$                  & $5.53$                  \\
\end{tabular}
\end{center}
\end{table}
In these tables only charge densities on the side of the sample closer to the piezoelectric sensor have been measured since we expect them to be more accurate \cite{hoang14}.

\section{Discussion}

According to our initial premises, uniform polarization, as obtained in experiments $A$ and $B$, should be difficult to detect in any way, either direct or indirect. In these cases, induced charge takes the form of a surface charge distribution at the internal face of the electrodes that has the same magnitude and opposite sign as the image charge in the electrodes once the sample is short-circuited. In a first analysis, the PEA signal from induced charge and image charge should cancel because they would have the same magnitude, opposite sign and their origins would be placed very close. Instead, image charge is very clearly seen in both electrodes of the $A$ and $B$ charge profiles of Figure~\ref{PEA-relax}. This situation has been observed previously on similar polymers in equivalent experiments~\cite{hoang14}.

If we compare the polarization of the relaxation measured by TSDC as presented in Table~\ref{polarization} with the charge per unit area presented in Table~\ref{sigma_density}, we can see that for experiment $A$ and $B$ the polarization is slightly lower but of the same order of magnitude than the image charge per unit area measured by PEA. This indicates that the experimental workflow is consistent and gives acceptable results.

\begin{figure}
\begin{center}
\includegraphics[width=12cm]{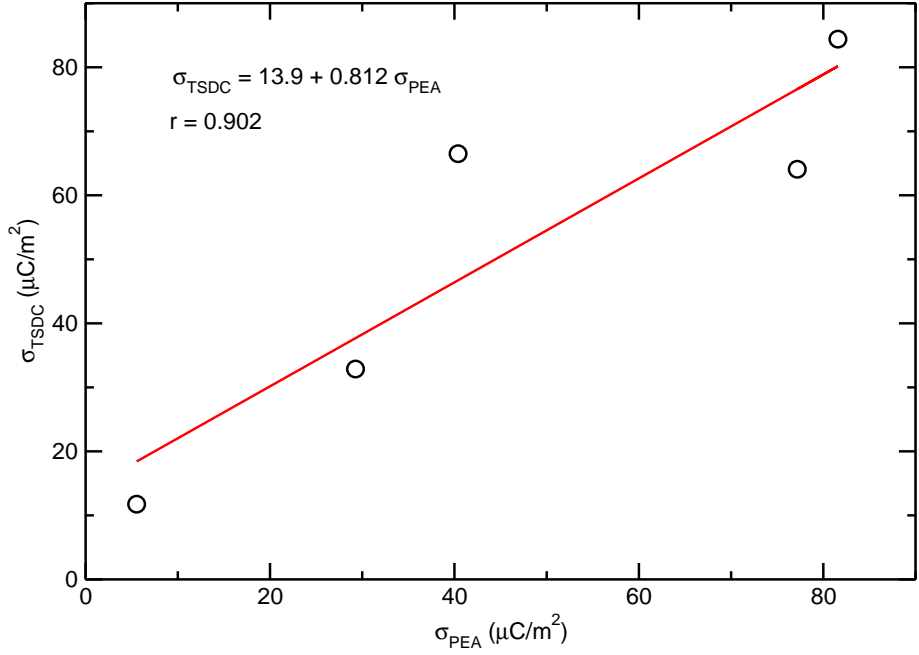}
\caption{Comparison of TSDC polarization and PEA image charge per unit area of samples with $\alpha$ relaxation activated.}\label{RMA2}
\end{center}
\end{figure}

We can take further this comparison plotting the data in Table~\ref{polarization2} in front of the data in Table~\ref{sigma_density2}. This plot, together with its linear regression, is presented in Figure~\ref{RMA2}.
The slope of the linear regression is still below but close to unity. Since the charge detected by TSDC is comparable to the image charge measured by PEA at the electrodes it follows that the detectability by PEA of induced charge related to the $\alpha$ and $\rho$ relaxations is really poor, if not none. About the discrepancy between these two values, it is difficult to assert if it is due to lack of efficiency the TSDC experiment or overestimation of charge during the PEA calibration process. 

With our current PEA equipment we can not discern the cause why the image charge on the electrodes gives a much more visible response than the induced charge in the sample, but we can make a plausible hypothesis to be tested with further means. Equation~\ref{pea_int} does not distinguish between the response of real charge in the electrodes (which is made of free carriers in a conductive medium) and the response of induced charge of an activated mechanism (which consists of bound charge in a dielectric medium). There may be a larger response time in the first case that gives a broader sound signal. This signal would be more easily detected by the piezoelectric sensor and its associated amplifier circuit. 

This discussion is based mainly on the data of the dipolar $\alpha$ relaxation because it is much easier to measure than the space charge $\rho$ relaxation. Anyway, data from the $B$ experiment supports that also in this case the charge profile consists mainly of image charge and induced charge is mostly unnoticed.

The profiles obtained from $C$, $D$ and $E$ are dominated by external charge injected at some depth in the bulk so we do not have to worry about the undetectability of induced charge, which should play a minor role. The broad profile and distance from the electrodes of external charge allows an easy detection with the PEA technique. 

The most striking feature of our results is that the qualitatively similar charge profiles $C$, $D$ and $E$ from Figure~\ref{PEA-80C} correspond to depolarization currents that have very different characteristics in terms of heterocurrents and homocurrents, as represented in Figure~\ref{TSDC-80C}. Depolarization current $C$ is heteropolar, $D$ has comparable heteropolar and homopolar parts while $E$ is predominantly homopolar. 

\begin{figure}
\begin{center}
\includegraphics[width=12cm]{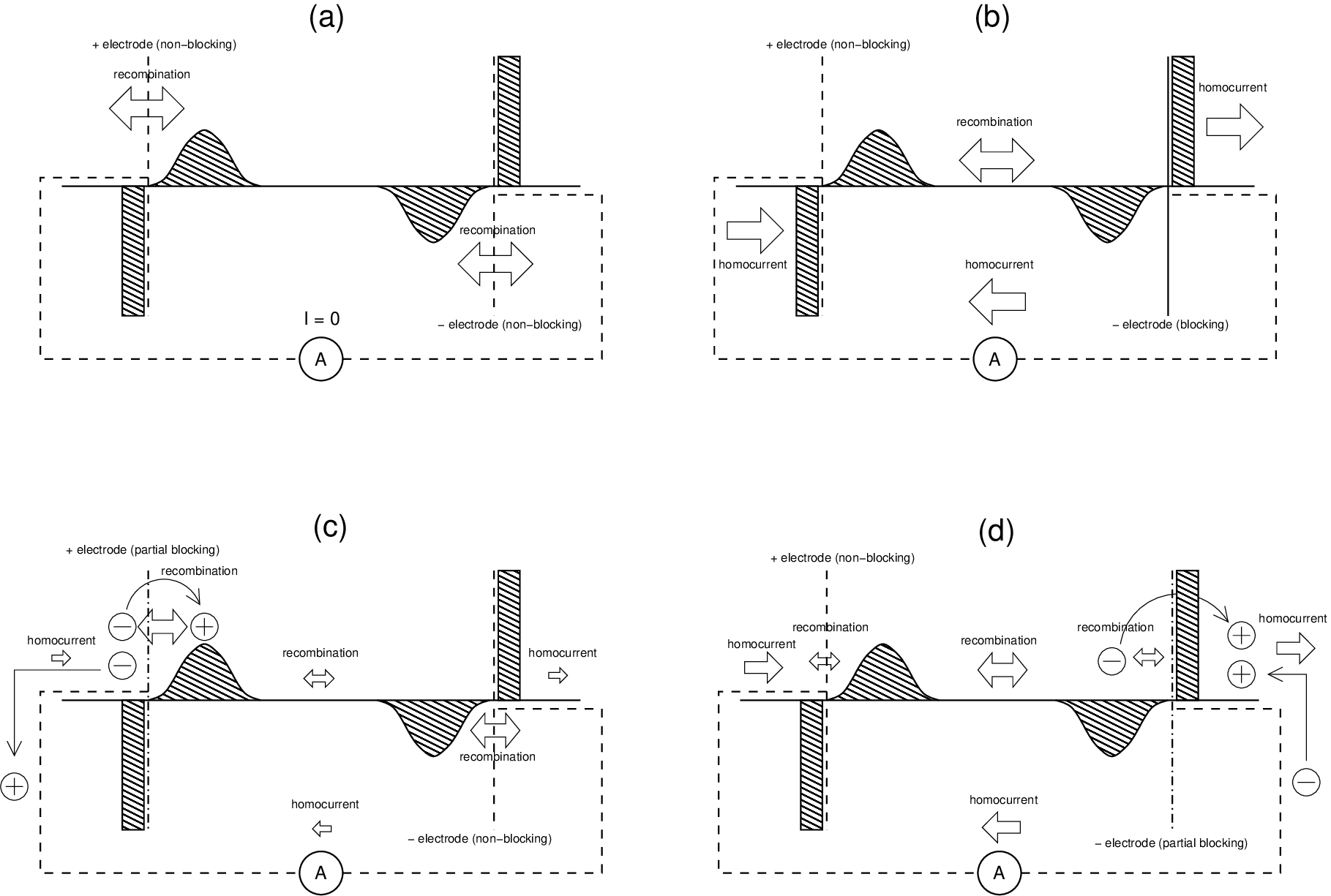}
\caption{Interpretation of the influence of the electrodes on TSDC results: (a) experiments $A$, $B$, $C$, (b) hypothetical experiment with fully blocking and non-blocking electrodes, (c) experiment $D$, (d) experiment $E$.}\label{interpretation}
\end{center}
\end{figure}

This paradox can be explained in terms of the blocking behavior of the electrodes. On the one hand, we can explain the lack of homocurrent in the depolarization current from $C$ if we assume that vacuum-deposited aluminum electrodes are non-blocking. As it is represented in Subfigure~\ref{interpretation}(a),
in this case recombination of external charge takes place predominantly at the electrodes and, as a result, no displacement current flows through the circuit. This situation would be completely opposite to the one represented in Subfigure~\ref{interpretation}(b), where no recombination can take place at the full blocking electrode and as a consequence it takes place entirely in the bulk of the sample. In this way a displacement homocurrent is registered which once integrated is theoretically equivalent to the image charge at one electrode.

On the other hand, the different magnitude of homocurrent in TSDC experiments $D$ and $E$ can be attributed to a different efficiency of blocking at the electrodes with an air gap in these experiments. In the same way that non-blocking behavior can be expected in vacuum-deposited aluminum electrodes, it is reasonable to assume that if there is a small air gap between the dielectric and the electrode then some amount of blocking can occur and an intermediate situation between the two aforementioned ones will arise. In Subfigure~\ref{interpretation}(c) we represent a case where the partially blocking electrode has a low blocking efficiency. Most of the recombination can still take place at the electrodes and the displacement current will be low, as observed in experiment $D$. Instead, as represented in Subfigure~\ref{interpretation}(d), if the partially blocking electrode has a higher efficiency then the recombination at the bulk will be reinforced and the displacement current will be higher, as happens in experiment $E$. 

The position of the partially blocking electrode in Subfigure~\ref{interpretation}(c) and~\ref{interpretation}(d) is also represented as in experiments $D$ and $E$, respectively, to have an appropriate representation of these experiments. 

Numerical analysis of the data in Tables~\ref{polarization} and~\ref{sigma_density} is consistent with these hypothesis. In experiment $C$ the polarization of the $\rho$ peak is one order of magnitude smaller than the image charge per unit area and, moreover, has opposite sign. This peak can not be related to external homocharge and has to be attributed to relaxation of uniform polarization, not registered by PEA. The external homocharge is mainly recombined at the electrodes without giving rise to homocurrent.

In experiment $E$ the polarization of the homocurrent peak is one order of magnitude greater than in experiment $D$ and very close to the value of the image charge per unit area. This is compatible with a highly blocking behavior of the air gap electrode in experiment $E$. In the case of experiment $D$ image charge per unit area values are comparable with $E$ but the polarization of the homocurrent peak $\rho^*$ is much lower, compatible with a less blocking behavior of the air gap electrode in this case.

From Subfigures~\ref{interpretation}(c) and~\ref{interpretation}(d) it can be deduced in which direction the charge carriers move in these experiments. In experiment $D$ electrons have to cross the partially blocking electrode from the electrode into the dielectric to recombine with the vacancies while in experiment $E$ electrons cross the partially blocking electrode from the dielectric into the electrode. 

Therefore, our results can be explained assuming that for electrons it is easier to cross the electrode--air--dielectric interface from the electrode into the dielectric than in the opposite way.

Even if numerical data obtained with TSDC and PEA support this interpretation, there remains the question of whether this behavior is consistent with current knowledge about this matter.   

It has been reported previously that vacuum-deposited electrodes present no initial potential barrier for electron injection or extraction because the Fermi level of the metal and its equivalent in the dielectric tend to level at the junction between both materials \cite{neagu08b}. So, it is reasonable to assume that vacuum-deposited electrodes are mostly non-blocking even in short-circuit conditions. 

With regards the electrodes with an air gap, the fact that there is not an intimate contact between the electrode and the dielectric can not only difficult the transfer of electrons (and, therefore, partially block the charge carrier flow) but it can also impede the leveling of the Fermi levels of the electrode and the dielectric. Since the work function is larger for PET than for typical metals (aluminum, steel, \ldots)~\cite{rajopadhye86} that would give a greater probability for an electron transfer from the electrode to the dielectric than the other way round. In any case, this is just an argument of plausibility that should be tested by further work. 

\section{Conclusions}

We have improved a simple but useful calibration scheme for PEA measurements on thin samples. It is exact while $C$ in Equation~\ref{pea_int} is not too different across the setup and sound wave transmission is complete across the interfaces. Since these conditions are not exactly met in usual cases there is room for further improvements, even at the expense of more complication in the calculations and the procedures.

One of the limitations of the calibration process is that the deconvolution matrix is obtained from a calibration case where there is only free charge. For this reason it can be expected to work better for image charges than for charges in the bulk of the sample.

Uniform polarization such as the one of $\alpha$ and $\rho$ relaxations can only be detected indirectly because induced and image charge are too close. In fact, the charge profile obtained corresponds mainly to image charge. The reason why image charge on the electrodes is much more visible than induced charge on the sample is unclear but may be related to different responses of both kinds of charge to the electric pulse combined with the frequency response of the piezoelectric sensor. The classical interpretation of these relaxations is reinforced because none of them appears to involve macroscopic displacement of charge.

Vacuum-deposited aluminum electrodes do not avoid charge injection under a strong electric field but under short-circuit they suppress homocurrent because they favor that recombination takes place at the electrodes without homocurrent flow through the circuit. This, is, nonetheless, useful in TSDC relaxational studies where often the interest is in the intrinsic mechanisms of the material rather than in the behavior of external charge.

In fact, electrodes are more determinant in the discharge of the sample than in its charge. None of the electrodes that have been used blocks charge carriers when a strong electric field is applied but differences appear when the sample is short-circuited, probably due to the interplay between electronic levels. Differences in the blocking behavior of the electrodes during the discharge trigger the appearance of homocurrent in TSDC experiments.

In electrodes with an air gap this blocking behavior depends on the direction of the flow of the charge carriers, changing the amount of homocurrent depending on which electrode has the air gap. We can infer that electrons move more easily in the aluminum-air-PET direction than in the opposite way.

Numerical analysis of TSDC and PEA data support these conclusions with a reasonable concordance between the polarization obtained by time integration of the depolarization currents and the charge per unit area obtained by integration along the depth axis of the charge profile.

PEA can always supply information about thermally poled electrets and its results can be successfully compared with TSDC measurements. External injected charge results are easier to interpret but mechanisms that give rise to a more uniform polarization can also be studied, albeit in an indirect way. In combination with techniques such as TSDC, PEA can be usefully employed in relaxational studies of dielectric materials, particularly in the identification of the different mechanisms that intervene in the response of the material.

\bibliographystyle{unsrt}
\bibliography{profile}

\begin{thebibliography}{10}

\bibitem{eguchi25}
Mototar{\^o} Eguchi.
\newblock On the permanent electret.
\newblock {\em The London, Edinburgh, and Dublin Philosophical Magazine and
  Journal of Science}, 49(289):178--192, 1925.

\bibitem{gerhard87}
R.~Gerhard-Multhaupt.
\newblock Electrets: Dielectrics with quasi-permanent charge or polarization.
\newblock {\em Electrical Insulation, IEEE Transactions on}, EI-22(5):531--554,
  Oct 1987.

\bibitem{chen81}
R.~Chen and Y.~Kirsh.
\newblock {\em Analysis of Thermally Stimulated Processes}, chapter~3, pages
  60--81.
\newblock Pergamon, Oxford, 1st edition, 1981.

\bibitem{vanturnhout99}
J.~van Turnhout.
\newblock {\em Electrets}, volume~1, chapter~3, pages 81--215.
\newblock Laplacian, Morgan Hill, {CA}, 3rd edition, 1999.

\bibitem{belana85}
J.~Belana, P.~Colomer, S.~Montserrat, and M.~Pujal.
\newblock Glass transition temperature of amorphous poly(ethylene
  terephthalate) by thermally stimulated currents.
\newblock {\em Macromol. Sci. Phys. B}, 23(4-6):467--481, 1984--1985.

\bibitem{mudarra97}
M~Mudarra and J~Belana.
\newblock Study of poly (methyl methacrylate) space charge relaxation by
  {TSDC}.
\newblock {\em Polymer}, 38(23):5815--5821, 1997.

\bibitem{montanari05}
G.C. Montanari and P.H.F. Morshuis.
\newblock Space charge phenomenology in polymeric insulating materials.
\newblock {\em Dielectrics and Electrical Insulation, IEEE Transactions on},
  12(4):754--767, Aug 2005.

\bibitem{ono04}
Ryo Ono, Masaaki Nakazawa, and Tetsuji Oda.
\newblock Charge storage in corona-charged polypropylene films analyzed by lipp
  and tsc methods.
\newblock {\em IEEE Transactions on Industry Applications}, 40(6):1482-- 1488,
  2004.

\bibitem{maeno88}
T.~Maeno, T.~Futami, H.~Kushibe, and T.~Takada.
\newblock Measurement of spatial charge distribution in thick dielectrics using
  the pulsed electroacoustic method.
\newblock {\em IEEE Transactions on Dielectrics and Electrical Insulation},
  23(3):433--439, 1988.

\bibitem{banda18}
ME~Banda, V~Griseri, Gilbert Teyssedre, and S{\'e}verine Le~Roy.
\newblock Polarization of electron-beam irradiated {LDPE} films: contribution
  to charge generation and transport.
\newblock {\em Journal of Physics D: Applied Physics}, 51(15):155303, 2018.

\bibitem{kazansky96}
P.~G. Kazansky, A.~R. Smith, P.~St.~J. Russell, G.~M. Yang, and G.~M. Sessler.
\newblock Thermally poled silica glass: Laser induced pressure pulse probe of
  charge distribution.
\newblock {\em Applied Physics Letters}, 68(2):269--271, 1996.

\bibitem{hoang14}
M-Q Hoang, L~Boudou, S~Le Roy, and G~Teyssedre.
\newblock Dissociating space charge processes from orientation polarization in
  poly(ethylene naphthalate) films.
\newblock {\em Journal of Physics D: Applied Physics}, 47(45):455306, 2014.

\bibitem{lewiner05}
Jacques Lewiner, S.~Hole, and Thierry Ditchi.
\newblock Pressure wave propagation methods: a rich history and a bright
  future.
\newblock {\em Dielectrics and Electrical Insulation, IEEE Transactions on},
  12(1):114--126, Feb 2005.

\bibitem{hole08}
S.~Hole.
\newblock Resolution of direct space charge distribution measurement methods.
\newblock {\em Dielectrics and Electrical Insulation, IEEE Transactions on},
  15(3):861--871, June 2008.

\bibitem{chen07}
G~Chen, Z~Xu, and L~W Zhang.
\newblock Measurement of the surface potential decay of corona-charged polymer
  films using the pulsed electroacoustic method.
\newblock {\em Measurement Science and Technology}, 18(5):1453, 2007.

\bibitem{sessler97}
G.~M. Sessler.
\newblock Charge distribution and transport in polymers.
\newblock {\em IEEE Transactions on Dielectrics and Electrical Insulation},
  4(5):614--628, 1997.

\bibitem{chen06}
G~Chen, Y~L Chong, and M~Fu.
\newblock Calibration of the pulsed electroacoustic technique in the presence
  of trapped charge.
\newblock {\em Measurement Science and Technology}, 17(7):1974--1980, 2006.

\bibitem{ahmed97}
N.~H. Ahmed and N.~N. Srinivas.
\newblock Review of space charge measurements in dielectrics.
\newblock {\em IEEE Transactions on Dielectrics and Electrical Insulation},
  4(5):644--656, 1997.

\bibitem{li94}
Ying Li, Masata ka~Yasuda, and Tatsuo Takada.
\newblock Pulsed electroacoustic method measurement of charge for accumulation
  in solid dielectrics.
\newblock {\em IEEE Transactions on Dielectrics and Electrical Insulation},
  1(2):188--195, 1994.

\bibitem{fleming05}
R.~J. Fleming.
\newblock Space charge profile measurement techniques: Recent advances and
  future directions.
\newblock {\em IEEE Transactions on Dielectrics and Electrical Insulation},
  12(5):967--978, 2005.

\bibitem{rizzo19}
Giuseppe Rizzo, Pietro Romano, Antonino Imburgia, and Guido Ala.
\newblock Review of the pea method for space charge measurements on hvdc cables
  and mini-cables.
\newblock {\em Energies}, 12(18), 2019.

\bibitem{sellares10}
J~Sellar\`es, JA~Diego, and J~Belana.
\newblock A study of the glass transition in the amorphous interlamellar phase
  of highly crystallized poly (ethylene terephthalate).
\newblock {\em Journal of Physics D: Applied Physics}, 43(36):365402, 2010.

\bibitem{arnaout16}
Mohamad Arnaout, Khaled Chahine, Fulbert Baudoin, Laurent Berquez, and Denis
  Payan.
\newblock Iterative deconvolution of pea measurements for enhancing the spatial
  resolution of charge profile in space polymers.
\newblock {\em Journal of Polymers}, 2016:4684796(11), 2016.

\bibitem{scipy}
Eric Jones, Travis Oliphant, Pearu Peterson, et~al.
\newblock {SciPy}: Open source scientific tools for {Python}, 2001--.

\bibitem{lide04}
David~R Lide.
\newblock {\em CRC handbook of chemistry and physics}, volume~84, chapter~13.
\newblock CRC press, 2003.

\bibitem{colomer98}
S.~Montserrat P.~Colomer and J.~Belana.
\newblock Relaxations in semicrystalline polyethyleneterephthalate using
  thermally stimulated currents.
\newblock {\em Journal of materials science}, 33(7):1921--1926, 1998.

\bibitem{neagu08b}
Eugen~R Neagu.
\newblock A method to measure the electric charge injected/extracted at the
  metal-dielectric interface.
\newblock {\em Applied Physics Letters}, 92(18):182904, 2008.

\bibitem{rajopadhye86}
NR~Rajopadhye and SV~Bhoraskar.
\newblock Ionization potential and work function measurements of pp, pet and
  fep using low-energy electron beam.
\newblock {\em Journal of materials science letters}, 5(6):603--605, 1986.

\end{thebibliography}

\end{document}